# *Quant Bust 2020*

## Zura Kakushadze[§†1]

§ *Quantigic® Solutions LLC,[2] 1127 High Ridge Road, #135, Stamford, CT 06905*
† *Free University of Tbilisi, Business School & School of Physics*
*240, David Agmashenebeli Alley, Tbilisi, 0159, Georgia*

April 7, 2020

*"You know, I couldn't do it.  I couldn't reduce it [why spin-1/2 particles obey Fermi-Dirac statistics] to the freshman level.  That means we really don't understand it."*
*– Richard Feynman[3]*

### Abstract

We explain in a nontechnical fashion why dollar-neutral quant trading strategies, such as equities Statistical Arbitrage, suffered substantial losses (drawdowns) during the COVID-19 market selloff.  We discuss: (i) why these strategies work during "normal" times; (ii) the market regimes when they work best; and (iii) their limitations and the reasons for why they "break" during extreme market events.  An accompanying appendix (with a link to freely accessible source code) includes backtests for various strategies, which put flesh on and illustrate the discussion in the main text.

---

[1] Zura Kakushadze, Ph.D., is the President and a Co-Founder of Quantigic® Solutions LLC and a Full Professor in the Business School and the School of Physics at Free University of Tbilisi. Email: zura@quantigic.com

[2] DISCLAIMER: This address is used by the corresponding author for no purpose other than to indicate his professional affiliation as is customary in publications. In particular, the contents of this paper are not intended as an investment, legal, tax or any other such advice, and in no way represent views of Quantigic® Solutions LLC, the website www.quantigic.com or any of their other affiliates.
[3] [Goodstein, 1989].

The recent coronavirus outbreak (the infectious disease COVID-19 caused by the virus SARS-CoV-2 [CDC, 2020]), that originated in Wuhan, China, has caused a stock market rout of the kind most young traders and other finance professionals have not seen before.  The fact that long positions were liquidated and trillions of dollars were wiped out in market capitalization is unsurprising.  However, what many apparently did find surprising is that, quant trading strategies, portfolios and funds, many of which are dollar-neutral (i.e., hold equal long and short positions), got obliterated in the process.[4]  And all this transpired despite most hedge funds nowadays – following the hype – using sophisticated machine learning (ML) and artificial intelligence (AI) techniques in identifying their portfolio composition as well as trading signals.

It is natural to wonder, how come?  Is it not the whole point of having a dollar-neutral portfolio not to be sensitive to market ups and downs?  A short answer is: not quite.  The goal of this note is to provide a longer answer in a relatively nontechnical fashion, by i) elaborating on why dollar-neutral strategies (such as statistical arbitrage, or StatArb) work during "normal" times in the first instance, and ii) thereby understanding their limitations and why they "break" during violent market selloffs (and not only).  It is precisely the lack of such understanding that trips up both fund managers and investors when they believe that dropouts with no requisite knowledge of the financial markets, whose skills are limited to knowing how to call off-the-shelf Python libraries,[5] will magically decipher the stock market by what not that long ago was called "data mining" (with a derogatory connotation) and now is artfully disguised as "data science".

So, what happened to quant strategies during what we term here as ***Quant Bust 2020*** triggered by the COVID-19 market selloff?  To address this question, let us focus on *dollar-neutral* quant trading strategies, as long-only strategies[6] generally are expected to lose money during market routs.  Furthermore, a priori dollar-neutral strategies can also vary substantially in their characteristics, including the holding period,[7] the number of stocks held in the portfolio, and how the portfolio positions are determined.  Here we will primarily focus on StatArb, which trades portfolios composed of a sizable number of stocks (typically, a few thousand most liquid U.S. stocks), with the portfolio composition and trading signals[8] determined algorithmically, based on some systematic quantitative approach (which can and often does include ML/AI techniques).  This is to be contrasted with the so-called long-short fundamental trading strategies, which trade many fewer stocks based on research done by human analysts and the trading decision-making includes substantial human-discretionary component.[9]  Furthermore, here we will focus on StatArb portfolios with "medium" holding periods, which very roughly can be stated as 1-20 days, albeit they can sometimes be somewhat longer or intraday.[10]  To be clear, we specifically exclude strategies with holding periods of months or years, as well as high frequency trading (HFT)[11] strategies operating on ultra-short horizons (milliseconds and below).

---

[4] For some media reports, see, e.g., [Aliaj, Wigglesworth and Kruppa, 2020], [Aliaj, Wigglesworth and Sender, 2020], [Burton, 2020], [Fletcher, 2020], [Teodorczuk, 2020], [Wigglesworth, 2020], [Wigglesworth and Aliaj, 2020]. Expectedly, there has also been a lively discussion in the social media, much of which, however, has been plagued with misconceptions and erroneous promulgations, which obscure the underlying reasons for this quant debacle.
[5] Without understanding how they work, or that they were not specifically developed for financial applications.
[6] Or, more generally, not just long-only but also net-long strategies, such as 130/30 – see, e.g., [Lo and Patel, 2008].
[7] Which is related to the daily portfolio turnover defined as an average fraction of the total portfolio dollar holdings traded (or "turned over", using the trader lingo) daily.  The turnover is inversely proportional to the holding period.
[8] Or alphas, using the trader lingo, which are essentially the expected (future) returns of the stocks in the portfolio.

To understand why StatArb works during "normal" times in the first instance, it is important to appreciate that the profitability of such quant trading strategies is due to a trickle-down effect stemming from long horizons.  At long holding horizons (measured in years) long-only investment strategies (mutual funds, pension funds, etc.) dominate.  Their portfolios are primarily based on fundamental considerations such as book-to-price (B/P) ratio, price-to-earnings (P/E) ratio, dividend yield, etc.[12]  This can make stock returns for different companies correlated if the latter are related by some pertinent similarity criterion, e.g., if they belong to the same sector, industry or sub-industry.  However, even if the stock price forecasts based on fundamental analyses were foolproof (which they are not as there is much uncertainty in them), in the market stock prices are determined by supply and demand.  Thus, due to a large number of market participants and the virtual impossibility to predict supply and demand imbalances or their precise timings, mispricings and inefficiencies are inevitable.  Furthermore, when a long-horizon long-only institutional player decides to buy a stock (or liquidate/reduce an existing position), it is not concerned with the price at which it wishes to execute the trade with a penny-order-of-magnitude precision, and rightfully so.  First, if the position is to be held for many months or years, it is virtually impossible to forecast the target future price with such precision.  Second, such entities execute large order sizes[13] and market impact is unavoidable.  So, worrying about pennies per share is impracticable.  In contrasts, quant trading strategies have much slimmer margins, where not only each penny but a tenth of a penny per share can make or break a strategy.[14]  So, long-horizon strategies create arbitrage opportunities on somewhat shorter horizons; strategies on such scales create arbitrage opportunities on yet shorter time scales; and so on – all the way down to HFT strategies.  This is the aforesaid trickle-down effect – the profitability of quant strategies (which can be dollar-neutral) ultimately comes from the long-horizon long-only strategies.  And the reason why quant strategies work is due to the aforesaid correlations between various stocks created by long-horizon strategies.

---

[9] See, e.g., [Jacobs and Levy, 1993].  In hybrid "quantamental" strategies quantitative tools are utilized (often along with human-discretionary research) for stock picking – see, e.g., [Slimmon and Delany, 2018].  Typically, such portfolios still consist of a modest number of stocks, as opposed to a few thousand stocks in StatArb portfolios.

[10] Even, say, 10- or 20-day holding period StatArb strategies have intraday components as they rebalance intraday.

[11] See, e.g., [Aldridge, 2013], [Jarrow and Protter, 2012], [Lewis, 2014], [O'Hara, 2015] and references therein.

[12] See, e.g., [Basu, 1977], [Black and Scholes, 1974], [Campbell and Shiller, 1998], [Chan, Jegadeesh and Lakonishok, 1996], [Fama and French, 1998], [Rosenberg, Reid and Lanstein, 1985], [Stattman, 1980] and references therein.

[13] Albeit, not necessarily at once; orders can be broken up into smaller sequential orders to reduce market impact.

[14] Quant strategies have much higher turnover than long-horizon ones, so it is a slim-margin/high-volume game.

So, inefficiencies lead to volatility and mispricings at various time horizons.[15]  These mispricings are then arbitraged away by what can be viewed as "mean-reversion" (or "contrarian") strategies.[16]  The premise here can be understood on the simplest mean-reversion strategy: pairs trading.[17]  This dollar-neutral strategy amounts to identifying a pair of historically highly correlated stocks and, when a mispricing (i.e., a deviation from the high historical correlation) occurs, shorting the "rich" stock (whose return is higher than the mean return of the two stocks) and buying the "cheap" stock (whose return is lower than said mean return).  This can also be done for multiple stocks (e.g., those in the same industry): rich/cheap (w.r.t. the industry-mean return) stocks are shorted/bought (within a dollar-neutral portfolio).

There is also a flipside.  For a given mean-reversion time horizon there may also exist opportunities to profit via what can be viewed as "momentum" strategies on somewhat shorter and/or longer time horizons (e.g., while the mispricing is occurring, say, when one of the stocks in a pair runs up compared with the other stock; or, e.g., if mean-reversion "overshoots" in correcting an existing mispricing thereby creating a new mispricing in the opposite direction).

In practice, while there are pure mean-reversion and pure momentum quant strategies, many strategies intertwine (often unwittingly) mean-reversion and momentum in ways that make it virtually impossible to detangle these components.  This is especially so in the case of uninterpretable black-box ML/AI based strategies, which often combine large numbers (millions and more) of weak trading signals such that the combined signal is strong and thus tradable.[18] In many cases quant trading strategies can even end up being factor investment strategies in disguise.[19]  So, an inexperienced quant can get easily confused (and confuse investors) with all this complexity as to why a quant strategy works or stops working under a given market regime.

However, to demystify things, let us focus on, say, dollar-neutral mean-reversion equity StatArb strategies trading a few thousand (e.g., 2,000-2,500) most liquid U.S. stocks in their portfolios, with 1-20 day holding periods (see above).  So, how is such a portfolio constructed?

In his seminal work on mutual fund performance,[20] Sharpe [1966] eloquently posits: "The key element in the portfolio analyst's view of the world is his emphasis on both expected return and risk."  Construction of a trading portfolio for equities StatArb schematically can be thought of as consisting of two steps. First, one comes up with some expected returns for stocks in the trading universe. This drives the "reward" part of the portfolio. Second, one constructs the portfolio holdings based on these expected returns. It is mainly this stage that deals with the "risk" part.[21] In the context of mean-reversion StatArb, a basic (but not the only) way of constructing the expected return for a given stock is, e.g., to average its daily close-to-close returns[22] over some number of prior days. So, e.g., a 10-day mean-reversion StatArb strategy in computing its trading signal for trading today would use a 10-day average daily close-to-close returns computed based on the 10 prior trading days' prices.[23] However, one still needs to know what these returns are mean-reverting against, i.e., the "benchmark" returns.

---

[15] For more on the trickle-down effect, see [Kakushadze, 2015d] (cf. [Lo, 2008]).  Also, once all players and costs are accounted for, the market is a "zero-sum" game [Kakushadze, 2015d] (cf. [Sharpe, 1991], [Pedersen, 2018]).
[16] For some literature on mean-reversion strategies, see, e.g., [Avellaneda and Lee, 2010], [Black and Litterman, 1991], [Idzorek, 2007], [Jegadeesh and Titman, 1995], [Kakushadze, 2015a], [Kakushadze and Serur, 2018], [Lakonishok, Shleifer and Vishny, 1994] and references therein.
[17] See, e.g., [Engle and Granger, 1987], [Gatev, Goetzmann and Rouwenhorst, 2006], [Kakushadze, 2015a].
[18] The literature on this subject is extremely scarce due to the highly secretive nature of this subject.  See, e.g., [Kakushadze, 2016b], [Kakushadze and Tulchinsky, 2016], [Kakushadze and Yu, 2017a], [Tulchinsky et al, 2015].
[19] On factor investing, see, e.g., [Blitz, 2015] and references therein.
[20] This paragraph (with some tweaks) is borrowed from [Kakushadze, 2016a].

A priori these "benchmark" returns can be anything, they do not even have to have a clear financial or economic interpretations. In fact, in black-box ML/AI strategies this is often the case. However, based on our discussion above, the correlations between various stocks that underlie mean-reversion StatArb are not random and stem from longer-horizon strategies. In this regard, there are essentially two types of fundamental (non-statistical/non-data-mined) "benchmark" returns – also known as factors (see below) – one can distinguish. First, there are the so-called style factors based on some estimated/measured properties of stocks. Examples of style factors are size (log of market capitalization), value (B/P ratio), growth (earnings growth), momentum (historical return), liquidity (average daily dollar volume, or ADDV), volatility (historical or option-implied standard deviation), etc.[24] Second, there are industry factors based on a similarity criterion such as stocks' membership in sectors, industries, sub-industries, etc. (depending on a particular nomenclature and desired granularity) under a given industry classification.[25] For long-horizon investment strategies with low turnover typically the granularity of the industry classification is limited to sectors (whose number is of order 10) or industries (whose number is of order 50). However, for short-horizon trading strategies with higher turnover such as mean-reversion StatArb, usually higher granularity at the level of sub-industries (or similar) is required, and their number for a typical trading universe is in hundreds. On the other hand, the number of style factors is only of order 10, and the number of relevant style factors is even smaller (4 or fewer) for short-horizon trading strategies.[26] As a result,

---

[21] Albeit, some elements of "risk management" can be (and often are) incorporated into the expected returns, e.g., sector/industry-neutrality (see below; albeit, in our examples industry-neutrality is incorporated at the "risk" step).

[22] By the daily close-to-close return (using the trader lingo) for a given trading day $T$ we mean the return from the previous trading day's price at the market close (which, with the exception of a few special days during the year, is 4 PM Eastern Time, when trading at the New York Stock Exchange closes) to the price at the market close on day $T$.

[23] I.e., going back 10 trading days starting with yesterday (or whatever the previous trading day may be), inclusive.

[24] See, e.g., [Kakushadze and Liew, 2015] for references.

[25] Such as GICS (Global Industry Classification Standard), BICS (Bloomberg Industry Classification System), SIC (Standard Industrial Classification), ICB (Industry Classification Benchmark), etc.

[26] See [Kakushadze and Liew, 2015], [Kakushadze, 2015b], [Kakushadze and Yu, 2016a]. In a nutshell, long-horizon style factors such as growth, which is based on multi-quarter earnings changes, are of no import for holding periods measured in days. This decoupling of time horizons was discussed in [Kakushadze and Liew, 2015]. To

purely due to their ubiquity, industry factors overwhelmingly dominate compared with style factors.[27] So, for our purposes we can focus on granular industry factors (sub-industries).[28]

There are many (in fact, infinite) ways of constructing a mean-reversion StatArb portfolio based on industry factors. However, the actual implementation will not be crucial for our purposes here. What is important is the concept. And the concept is that, we have factor returns for each industry,[29] and the returns for stocks belonging to a given industry are assumed to be mean-reverting around the factor return for said industry. As above, the rich stocks (w.r.t. to said return) are shorted and the cheap stocks are bought. A simple (but, once again, not the only) way to implement such a strategy is via a weighted cross-sectional regression. Such a portfolio is industry-neutral by construction.[30] Another construction is via mean-variance optimization[31] using a multifactor risk model which is based on (or includes) industry factors. In this case the portfolio is not industry-neutral[32] and the dollar-neutrality is imposed as a constraint in optimization. Either way, the premise behind mean-reversion is that stocks in the same industries historically are highly correlated and when these correlations are temporarily undone, they are expected to snap back to normal, hence arbitrage opportunities.

Now, this premise (largely) holds during "normal" times (see below). However, it breaks badly during violent market routs (when volatility spikes) such as the COVID-19 selloff. Why? The reason is quite prosaic. During a selloff both long-only and dollar-neutral portfolios are liquidated. However, long-only portfolios dominate by orders of magnitude, and here we are most physicists who have studied quantum field theory (QFT) this result would be almost axiomatic due to the well-familiar decoupling theorems in QFT (see, e.g., [Appelquist and Carazzone, 1975], [Symanzik, 1970], [Weinberg, 1973]). However, apparently, to most people in (quant) finance this is not at all evident without empirical proof, which is provided in [Kakushadze and Liew, 2015] along with intuitive theoretical considerations.

[27] See [Kakushadze and Yu, 2016a] for a detailed discussion and empirical analysis. Also see [Hong, Torous and Valkanov, 2007] for an empirical analysis in the context of longer horizons with essentially similar conclusions.

[28] To be clear, this is not to say that, e.g., earnings announcements are not important in short-horizon strategies, in fact, they are as volatility increases around earnings announcements. However, monitoring earnings announcements and, e.g., not trading stocks immediately following their earnings announcements to avoid getting tripped up by the increased volatility (see below) is not the same as using, e.g., a growth style factor (which, as mentioned above, is based on multi-quarter earnings changes) for, say, 10-day mean-reversion as this would only add noise to the mean-reversion signal, thereby leading to overtrading and unnecessarily increasing trading costs.

[29] Again, in most cases "industry" stands for "sub-industry" or a similar granular level in an industry classification.

[30] Stock returns are cross-sectionally regressed against a "dummy" (i.e., binary) matrix of 1s and 0s, whose rows are labeled by stocks and columns are labeled by industries, and whose element is 1 if a given stock belongs to a given industry, and 0 otherwise. The regression weights can be taken to be inverse historical variances of stock returns. This results in a portfolio which is automatically industry-neutral (i.e., for each industry the holdings are dollar-neutral, so the entire portfolio is also dollar-neutral). Furthermore, the stock holdings are suppressed by the aforesaid historical variances, so the contributions of the volatile stocks are suppressed. See [Kakushadze, 2015a].

[31] See [Markowitz, 1952]. Alternatively, the Sharpe ratio [Sharpe, 1994] can be maximized [Kakushadze, 2015a].

[32] The portfolio can be thought of as approximately industry-neutral (see [Kakushadze, 2015a]). In fact, it can be decomposed into a linear combination of two portfolios. The first portfolio, as above, is a dollar-neutral mean-reversion portfolio based on a weighted linear regression, so it is industry-neutral. The second portfolio is dollar-neutral and has exposures to industry factors (some long and some short), so it is a factor-momentum portfolio.

further interested in quant strategies.  So, while the following numbers are not meant to precisely reflect the proportions of long-only investment strategies vs. dollar-neutral quant trading strategies, they illustrate the key point here.  Thus, according to BarclayHedge,[33] in 2019, Q4 the total assets under management (AUM) for hedge funds[34] was $3.194T.  However, AUM of "equity market neutral" funds was $76.2B, and AUM of "equity long/short" funds was $196.3B.  Some quant dollar-neutral trading strategies may also be grouped under the "multi-strategy" funds, whose total AUM was $360.7B.[35]  On the other hand, according to YCharts,[36] the U.S. total market capitalization peaked at 158.9% of GDP (gross domestic product) on February 19, 2020, and bottomed out at 103.4% of GDP on March 23, 2020 (resulting in a peak-to-trough drawdown of 34.93%).[37]  Using the 2019, Q4 GDP level of $21.73T [BEA, 2020], we thus get roughly $12T for the peak-to-trough drawdown in the U.S. total market capitalization.

So, the selloff was driven by the long-only liquidations.  But, again, the question is, why does this affect dollar-neutral quant trading strategies so adversely?  The answer is that, when large long-only liquidations begin, they do not preserve the "normal" correlations structure between, e.g., stocks in the same (sub-)industries.  Metaphorically speaking, such liquidations do not proceed in an "orderly fashion".  Indeed, large long-only liquidations arise by selling, among other things, large index-linked funds (mutual funds and ETFs) and futures.  Some of these indexes are market capitalization weighted, some have other weighting schemes.  Either way, there is absolutely no reason why the aforesaid "normal" correlation structure would be preserved in such liquidations and, in fact, it is not.  Instead, most correlations go out of whack.

Now, mean-reversion StatArb is a reactive strategy.  When a mispricing (as compared with its normal, historical expectation) occurs, it shorts what it perceives to be rich stocks and buys cheap stocks – anticipating a snap-back.  However, during extreme market routs there is no snap-back.[38]  The correlations become more and more undone as a result of long-only liquidations, so a StatArb strategy loads up more and more on losing positions.  This is basically a serial double-down on losing bets.  At some point the strategy loses money beyond its risk-tolerance parameters (especially when the positions are leveraged), so the portfolio manager starts to unwind it (typically, haphazardly).  These dollar-neutral liquidations compound on top of the long-only liquidations that triggered the debacle in the first instance further contributing into undoing the aforesaid "normal" correlation structure.  It becomes a self-fulfilling prophecy.

---

[33] See https://www.barclayhedge.com/solutions/assets-under-management/hedge-fund-assets-under-management/.
[34] Excluding funds of funds, whose total AUM during the same period was $290.4B.
[35] And also under "other" funds, whose total AUM was $43.2B.
[36] See https://ycharts.com/indicators/us_total_market_capitalization.
[37] On February 19, 2020 S&P500 was at 3,380.39 and peaked a tiny bit higher on February 20, 2020 at 3,380.45.  It bottomed at 2,290.71 on March 23, 2020.  So the peak-to-trough drawdown was 32.24% (consistently with the above YCharts number).  Historical data source: https://finance.yahoo.com/quote/%5EGSPC/history?p=%5EGSPC.
[38] At least for some time, until the market starts to bottom out, and the correlations begin to stabilize/normalize.

However, there is another effect that exacerbates the avalanche-like effect discussed above. On short time horizons it is precisely the StatArb strategies that keep the correlation structure intact. With StatArb liquidations the correlation structure goes haywire and all hell breaks loose. And what makes it worse is that mean-reversion StatArb is, by design, a liquidity-providing strategy. In the market regime where StatArb is expected to work well (see below), in the zeroth approximation mean-reversion StatArb is essentially a market-making strategy that attempts to fill its buy orders at the bid and sell-short orders at the ask.[39] During the market routs liquidity for sell-short orders dries up, so to maintain dollar-neutrality, the strategy must start executing more aggressively on that side[40] thereby getting execution prices that are much worse than the prices at which the strategy would normally expect to get orders filled.[41] This is exacerbated even further by the fact that under such conditions many stocks may become hard-to-borrow,[42] so some stocks become impossible or expensive (due to high stock borrow fees) to sell short.[43] On the other hand, the market trades right through the buy orders. So, the strategy is losing money not only because it is reactively doubling-down on losing trades, but also on both sides (buys and short-sales) of trade executions, so this is a double-whammy.

So, mean-reversion StatArb is simply not designed to make money or even fare well through rough market routs. What about rallying markets? When the market goes up, up and up on a straight-line trajectory, mean-reversion StatArb does not do all that well either. This is because there are not enough good arbitrage opportunities in such a market. While there is a clear asymmetry between rallies and selloffs,[44] during rallies long-only strategies also pump money, among other things, into large index-linked funds and futures, which is not necessarily conducive to creating "healthy" mean-reversion arbitrage opportunities. In such environments StatArb tends to underperform the market. Thus, S&P500 opened at 2,476.96 on January 2, 2019 and closed at 3,230.78 on December 31, 2019,[45] which translates into a 30.43% annual return for 2019. On the other hand, in 2019, BarclayHedge's Equity Market Neutral Index (based on 21 funds) had a negative return of -0.56%, Equity Long/Short Index (based on 29 funds) had a positive return of 6.59%, and Multi Strategy Index (based on 36 funds) had a positive return of 5.23%, all significantly underperforming S&P500.[46] In fact, when the market rallies, many dollar-neutral strategies that make money have a significant positive market beta,[47] either intentionally, or inadvertently, when alphas (trading signals) are found through extensive data mining, often using uninterpretable, convoluted black-box ML/AI algorithms.[48]

---

[39] This is not always possible and in real life more sophisticated executions are used [Kakushadze and Serur, 2018].

[40] Including sending out market orders or marketable limit orders. See [Kakushadze and Serur, 2018] for details.

[41] A.k.a. "slippage" (trader lingo); the buy orders (see below) get "adverse selection" [Kakushadze and Serur, 2018].

[42] See, e.g., [Engelberg, Reed and Ringgenberg, 2018] and references therein. The SEC (U.S. Securities and Exchange Commission) "Regulation SHO requires a broker-dealer to have reasonable grounds to believe that the security can be borrowed so that it can be delivered on the date delivery is due before effecting a short sale order in any equity security. This "locate" must be made and documented prior to effecting the short sale." [SEC, 2015].

[43] During 2002-2007, among other things, I was doing equities StatArb trading at RBC Capital Markets (RBCCM). At that time RBCCM had the largest Index Arbitrage (a.k.a. cash-and-carry arbitrage) book on Wall Street. An IndexArb book carries a long position in a diverse basket of stocks and an offsetting short position in index futures contracts (see, e.g., [Kakushadze and Serur, 2018]). Our dollar-neutral StatArb book consisted of long and "short" stock positions. However, thanks to our Index Arbitrage long stock holdings, our StatArb strategies almost never had to sell many of the stocks in their trading universe short. This gave us a sizable advantage over other players.

[44] During rallies money flows into the market so the overall "pie" increases, while during selloffs money flows out so the "pie" shrinks. Also, during rallies the liquidity issues with short-sales arising during selloffs largely are moot.

[45] Source: https://finance.yahoo.com/quote/%5EGSPC/history?p=%5EGSPC.

So, what is the market regime where mean-reversion StatArb works well? The answer jumps out from our discussion above: it is the so-called "sideways" market,[49] with decent volatility, but not too high. A sideways market is ideal as snap-back (mean-reversion) is pretty much guaranteed. Very low volatility is not conducive to generating arbitrage opportunities. On the other hand, a sideways market with too high volatility (a.k.a. "choppy" market) can have liquidity issues (large slippage), so the profitability of mean-reversion StatArb trading can suffer.

Based on the foregoing, it is clear that dollar-neutrality of quant trading strategies does not make them "market-neutral".[50] A mean-reversion StatArb, *effectively*, is akin to selling deep out-of-the-money put options on a broad market index. When a black-swan event[51] hits, such as the 2007 Quant Crash (a.k.a. 2007 Quant Meltdown) in August of that year,[52] or COVID-19,[53] there is just no way around it – this kind of strategies are *designed* to lose money. Any promulgations to the contrary are just smoke and mirrors. It is like claiming that an insurance company would not lose money as a result of a 9/11-like event. And no "bells and whistles" can change that fundamental fact. No fancy/convoluted ML/AI black-box could have averted Quant Bust 2020. Indeed, if your ML/AI could predict the COVID-19 market selloff, why would you want to trade StatArb?[54] You could simply (effectively) short the market (including, effectively with leverage) and make a huge return on capital (cf. Bill Ackerman's almost 100-fold return).[55] Put differently, StatArb and other quant trading strategies are not "all-weather" strategies. In fact, no strategy is "all-weather", or else multi-strategy hedge funds, which diversify across many different types of strategies, would not exist.[56]  No one complained about StatArb when it made high annualized Sharpe ratio (6+) returns with a few (or no) down months.  Now it lost money because it is *designed* to lose money in such events; it would be surprising if it did not.[57]

It is a separate issue that many investors (as well as some traders) do not understand idiosyncratic facts about StatArb and quant trading more generally.  The ML/AI hype did not help in this regard.  In fact, one can wonder, if ML/AI, which most (if not all) quant trading shops use these days, is such a "panacea", why did it not prevent Quant Bust 2020?  A simple answer is that ML/AI is not a panacea.  It cannot predict black-swan events any better than a simple linear regression can.  In fact, a sensationalist – and in this day and age sensationalism appears to have become a norm – could turn the tables around and blame ML/AI for Quant Bust 2020.  In fact, there are ways that ML/AI can make things worse in situations such as Quant Bust 2020.  How come?  ML (and especially AI)[58] can have a large number of often obscure (uninterpretable) parameters.  In various other applications (such as computer vision, document classification, some bioinformatics/computational biology problems, etc.) these parameters can in fact be stable out-of-sample.[59]  However, in trading applications such out-of-sample stability is rarely (if ever) the case for long periods of time, so trading signals tend to be ephemeral by nature.  Another way of phrasing this is that, in trading applications signal-to-noise ratios are notoriously low, and uncerebral, brute-force applications of ML/AI to such problems (all the hype notwithstanding) is rarely successful.  A much more successful approach is to carefully identify the noise and factor it out before applying ML/AI to discern the signal from so-denoised data – after all, no matter the number of computational cycles, by definition, *one cannot machine learn the noise*.  However, identifying and factoring out the noise is what requires deep understanding of the financial markets, which (quoting from above) "dropouts with no requisite knowledge of the financial markets, whose skills are limited to knowing how to call off-the-shelf Python libraries" do not possess, also by definition.  As is aptly put in [AQR, 2019], "finance is more complex than many other domains of ML research (cats don't begin morphing into dogs once the algorithm becomes good at cat recognition)."  Taking this analogy further, most ML/AI algorithms trained on cats and dogs would misidentify hyenas as canines, even though hyenas are much more closely related to cats: hyenas and cats belong to suborder *feliformia*, while dogs belong to suborder *caniformia* (both are suborders of order *carnivora*).

---

[56] To be clear, multi-strategy funds are not a free lunch.  They cost a lot of money to run and in good times employ hundreds of (or even more) people.  In contrast, Berkshire Hathaway's Omaha, Nebraska corporate office employs only 26 people (albeit its subsidiaries' employee count totaled over 391,000 in 2019) [Berkshire Hathaway, 2019].

[57] This exposure to market selloffs can in principle be hedged by buying deep out-of-the-money put options on a broad market index.  However, this costs money and cuts into the strategy returns.  Some studies argue (albeit in other contexts) against using protective puts [Israelov and Nielsen, 2015], [Israelov, Nielsen and Villalon, 2017].

[58] AI algorithms use large weights matrices (determined via training), which can easily have 1,000s of elements.

[59] In the ML/AI lingo, if this is the case, an algorithm generalizes well, i.e., the parameters discerned from a training dataset also work well on new datasets.  However, in many cases there can be overfitting of an algorithm, so it lacks predictive power when it comes to new datasets.  This can happen not only due to a badly built model, but also if the problem intrinsically has highly heterogeneous data and lacks generalizability from dataset to dataset.

So, if the underlying fundamental reason for why mean-reversion StatArb works during the "normal" times is rooted in long-horizon long-only investment strategies (see above), then what does ML/AI do in this context? There are (at least) two parts to it. First, there is the hype (and related smoke and mirrors). There is not much to add to that. However, there are ML/AI based trading signals that do work. In fact, there is a plethora of such trading signals. However, typically they are weak and short-lived, so to generate a strong trading signal, many (sometimes millions or more) such signals are combined with nontrivial weights (see above). However, one can still inquire, what is the underlying financial interpretation of these data-mined (however ephemeral and weak) ML/AI trading signals? The answer basically lies in the aforesaid trickle-down effect from long horizons to shorter ones: due to a large number of market participants and the virtual impossibility to predict supply and demand imbalances or their precise timings, mispricings and inefficiencies are inevitable and arise at various time horizons. Data-mining techniques can uncover the aforesaid ephemeral and weak trading signals, albeit in many cases it is virtually impossible to pinpoint a precise economic/financial reason for why a given such signal exists in the first instance.[60] It is then unsurprising that trading strategies based on such signals can stop working when a dramatic regime change (selloff) transpires in the marketplace.

Also, to be clear, not all quant trading strategies lost money during Quant Bust 2020. While mean-reversion StatArb – expectedly – got obliterated, other types of strategies did not, in fact, some even made money. For instance, momentum/trend following strategies that got the direction of the market right (be it due to sheer luck or some honest-to-goodness secret sauce) and went short would make money, but then again, such strategies would not be dollar-neutral. Strategies that had (intentional or unwitting) factor exposure could also make money, including dollar-neutral strategies. For instance, if a strategy had a substantial short exposure to oil stocks (which means that it was not industry-neutral or even approximately industry-neutral), then it might very well have made money during the COVID-19 selloff: WTI (West Texas Intermediate, a.k.a. Texas light sweet) crude oil price was trading at $53.86 a barrel on 2/20/2020 (the day S&P500 peaked at 3,380.45 – see above) and crashed to the low twenties in March 2020 (it was trading at $23.36 a barrel on March 23, 2020, the day S&P500 bottomed out at 2,290.71 – see above).[61] Those lucky (or smart) to be short oil stocks likely made money.

So, Quant Bust 2020 happened for a good reason, and it was not because ML/AI (or alternative data – see below) was not used. Quite the contrary, often the uninterpretability of ML/AI is what obscures things, for (see the Feynman quote above) if it cannot be reduced to the freshman level, it means that one does not understand it. Also, once again, ML/AI cannot predict black-swan events such as COVID-19. We know this to be true empirically for, if the opposite were the case, we would not be talking about Quant Bust 2020 (or Bill Ackerman's human-discretionary bet – see above). However, there is a prosaic theoretical reason for this as well. ML/AI needs to be trained on data to see patterns in new data.[62] If the patterns in new data are not contained in the training data, it is a tall order to expect an ML/AI algorithm to magically identify such patterns in new data.[63] As the 12th century Georgian poet Shota Rustaveli aptly put it, "What is in the jar is what will flow out."[64] It still holds nine centuries on.

---

[60] To be clear, these signals need not be purely "mean-reversion" or "momentum" signals. They can be convoluted and utterly indiscernible combinations of mean-reversion and momentum signals from a number of time horizons.
[61] Thus, from February 20, 2020 to March 23, 2020 oil lost 53.63% (and this rout was triggered by the Oil War [Cho et al, 2020]). Oil prices data source: https://www.macrotrends.net/2516/wti-crude-oil-prices-10-year-daily-chart.

To summarize, quant trading strategies are not "all-weather" or a free-lunch. As with everything else, they have limitations. *Así es la vida*. I hope the above discussion helps clear things up in this regard. Appendix A contains explicit backtests of mean-reversion (and other) trading strategies to put some flesh on and illustrate what happened during Quant Bust 2020. However, a word of caution: these backtests are by no means intended to be exhaustive in any sense, only illustrative. Not all quant trading strategies lost money amid the COVID-19 selloff.[65]

## Appendix A

StatArb strategies (except purely intraday ones) usually hold positions overnight.[66] However, overnight holdings increase portfolio volatility and can obscure the trading signal (alpha). So, here we "strip down" StatArb strategies such that they (at least on paper – see below) only have intraday holdings: they establish at the open and liquidate at the close of each trading day.  Now, this is not how things are done in practice,[67] but it suits our purposes here as we wish to illustrate how StatArb alphas went bad during Quant Bust 2020, even before any trading costs, due to the "normal" correlation structure between the stocks becoming undone.

---

[62] In this regard, one may wonder why alternative data (geolocation, satellite imagery, social media sentiment, credit card transactions, online browsing activity and other data exhaust, web-scraped job postings, flight and shipping trackers, etc. – see, e.g., [Drenik, 2019], [McPartland, 2017], [Monk, Prins and Rook, 2019]) did not make a big difference in the context of foreseeing the COVID-19 selloff (or, for that matter, why more generally it has not had the impact that might have been expected from it based on a prior hype). There is a simple explanation to this. Trading signals based on each individual alternative datafeed tend to be rather weak (and in many cases also ephemeral). So, to get a strong signal one would need to combine many such weak signals. However, there are not that many alternative data feeds to begin with. Thus, according to https://alternativedata.org/stats/, there are only 445 alternative data providers (as of April 3, 2020), and not all those sources necessarily add value in trading.
[63] To be clear, there are plenty of past occurrences where, say, mean-reversion StatArb lost money in high-volatility environments. Experienced quant traders know this and no ML/AI is needed to confirm this for the reasons discussed above. However, once again, mean-reversion StatArb is a reactive strategy and loses money in such environments by design. Training ML/AI on the past mean-reversion StatArb meltdowns is not going to make it "learn" anything new in this regard. It certainly will not make it any better at forecasting new black-swan events.
[64] This is my own translation of an aphorism (which, sadly, greatly suffers in translation) from a stanza in Rustaveli's sole known epic poem, whose title often is erroneously translated as "The Knight in the Panther's Skin" (or similar) – see, e.g., [Rustaveli, 2015]. In my humble opinion, not only is this poem the greatest masterpiece of the (rich) Georgian literature, but also one of the greatest literary writings of all time worldwide. It consists of over 1,600 perfectly rhymed *shairi* or Rustavelian quatrains, each of which contains 16 = 8 + 8 syllables per line with a caesura between the 8th and 9th syllables. How a human brain can come up with such perfection is mind-boggling, especially considering that this poem tells an extremely complex story complete with dialogs, aphorisms, etc.
[65] Our backtests in Appendix A suggest that, e.g., momentum strategies might have done well during that period.
[66] Albeit there are purely intraday StatArb strategies frequently trading in and out of positions throughout the day.

The backtrsting framework is discussed in detail in Appendix A of [Kakushadze and Serur, 2018], which also provides the backtesting source code.  Therefore, we will not rehash it here and instead will focus on the salient points.  The pricing data for the universe of all exchange-listed U.S. equities (common stocks and class shares only; no preferred shares, warrants, rights, ETFs, etc.) was downloaded from https://finance.yahoo.com on March 25, 2020 (so the data contains prices going back from March 24, 2020, inclusive).  However, since some of the backtests involve overnight returns (close-to-open returns), the most recent date for which P&L (profits and losses) is computed is March 23, 2020.  Also, the backtest goes back 1 year (252 trading days).[68]  The trading universe consists of 2,000 most liquid stocks by ADDV (average daily dollar volume) computed based on the prior 21 trading days (i.e., 1 month).  However, the trading universe itself is only updated every 21 trading days (so the backtesting period of 252 trading days consists of 12 such 21-trading-day periods, and the universe is updated only at the beginning of such 21-trading-day period).  The trading portfolio is always dollar-neutral.  Furthermore, the absolute[69] dollar holdings for each stock cannot exceed 1% of ADDV for that stock.  This ensures that the portfolio never trades more than 1% of ADDV for any given stock.  These trading bounds coincide with the position bounds in these strategies.

We backtest a number of different strategies.  There are 6 different types of returns used, 4 of which are 1-day, 5-day, 10-day and 20-day close-to-close returns (see above).  These returns are used in mean-reversion strategies.[70]  Another return is the previous trading day's open-to-close return,[71] and in this case we actually have a momentum strategy.  I.e., it bets not on mean-reversion (w.r.t. an industry-average return), but in the opposite direction (that the excess return w.r.t. said industry-average return on the previous trading day will continue to the next).  Finally, we have overnight close-to-open returns[72] used in mean-reversion strategies.

The second differentiating feature of the strategies is how the portfolios are constructed based on the aforesaid returns.  Following our discussion above, we use a weighted regression and mean-variance optimization, both with bounds (1% of ADDV on individual stock positions – see above).  Without bounds the weighted regression would be done as follows – albeit, to be clear, we always use bounds (see above and below).  Given an industry classification (see below) we have a "dummy" (i.e., binary) matrix of 1s and 0s, whose rows are labeled by stocks and columns are labeled by industries, and whose element is 1 if a given stock belongs to a given industry, and 0 otherwise.  We cross-sectionally regress the stock returns (see above) against this "dummy" matrix with the regression weighs given by inverse historical variances of daily close-to-close returns computed over the previous 21 trading days (1 month) and updated with the same 21-trading-day frequency as the trading universe (see above).  The stock dollar holdings are given by the negated residuals of the aforesaid cross-sectional weighted regression divided by the aforesaid historical variances.  The so-obtained stock holdings by construction are industry-neutral, and the entire portfolio is dollar-neutral.  With bounds things get trickier and we use the algorithm and source code for the bounded regression given in Appendix A of [Kakushadze, 2015c].  For mean-variance optimization with the dollar-neutrality constraint and bounds (1% of ADDV on individual stock positions – see above) we use the algorithm and source for the optimizer with constraints and bounds given in Appendix C of [Kakushadze, 2015e].  Mean-variance optimization also requires a multifactor risk model as an input.  We use heterotic risk models of [Kakushadze, 2015e] (also see [Kakushadze and Yu, 2016a])[73] based on a given industry classification (see below).  A given risk model is recomputed 100% out-of-sample with the same 21-trading-day frequency as the trading universe (see above).  The stock dollar holdings in the portfolio are dollar-neutral by construction (due to the dollar-neutrality constraint explicitly included in the optimizer), but are only approximately industry-neutral.

We use two types of industry classification.  One is based on fundamental data.  Above we discussed examples of such industry classifications (GICS, BICS, SIC, ICB, etc.).  Most of this data is not free and in fact is not cheap.  However, SIC (Standard Industrial Classification) data is available from the SEC website, albeit in a not-user-friendly format, including no up-to-date data for stock tickers (only company names).  Happily, open-source code for downloading and processing this data into a useful (for backtesting and other purposes) form is freely available in [Kakushadze and Yu, 2017b], so we use the most granular level of SIC (called "industries" in the SIC nomenclature, and essentially corresponding to sub-industries in GICS and BICS).  There are and Serur, 2018]): it uses today's open price to compute the trading signal to trade at today's open.  In real life there would always be a delay (however small), so if today's open price is used to compute the signal, then the strategy would have to execute the trades sometime after today's open.  So, this "delay-0" signal essentially measures the pure overnight alpha.  The actual real-life trading signal would be much weaker by sizable slippage.
[73] Source code for constructing heterotic risk models is given in Appendix B of [Kakushadze, 2015e].  Source code for a more general construction (which includes heterotic risk models as a particular case), which allows to incorporate style and principal component factors, is further given in Appendix A of [Kakushadze and Yu, 2016a].

roughly 320-325 SIC industries for a typical trading universe of 2,000 most liquid stocks (by ADDV) we use in our backtests (see above), so it is quite granular.[74]  Another type of "industry classification" we use does not involve any fundamental data, only the pricing data, is purely statistical, and utilizes unsupervised ML (clustering) techniques.  Source code for building this so-called "statistical industry classification" is given in Appendix A.2 of [Kakushadze and Yu, 2016b], which contains a detailed description of the clustering techniques used, rationale, etc.[75]

Finally, we run backtests both without (pure alpha) and with (simulated) trading costs. The costs are estimated exactly as in Appendix A of [Kakushadze and Serur, 2018] and on average correspond to assuming that, to establish/liquidate an equally-weighted portfolio, it costs 10 bps[76] of the total dollar amount executed.[77]  However, the cost model we use is not uniform across all stocks but takes into account their liquidity (ADDV) and historical volatility.[78]

The results of our backtests are summarized in Table 1 and Figures 1-6.  In Table 1 the legends are as follows.  The second column shows which industry classification is used: the legend "SIC" stands for the Standard Industrial Classification (based on fundamental data), and the legend "STAT" stands for the statistical industry classification (based on ML/clustering techniques) of [Kakushadze and Yu, 2016b] (see above).  The third column shows whether mean-variance optimization with bounds and the dollar-neutrality constraint (the legend "OPT") or the bounded weighted regression (the legend "REG") is used in the portfolio construction (see above).  The fourth column shows whether trading cost are included (the "Y" or "N" legends) in the backtest (see above).  The first column shows which return is used with the following legends: C2C1 = 1-day close-to-close return, C2C5 = 5-day close-to-close return, C2C10 = 10-day close-to-close return, C2C20 = 20-day close-to-close return (all mean-reversion strategies); MOM1 = previous trading days' open-to-close return (momentum strategy); D0 = previous trading days' close to current trading days' open (overnight) return (delay-0 mean-reversion strategy); see above for details.  The fifth column shows the annualized return-on-capital (ROC), which is computed as the average daily P&L multiplied by 252 (the number of trading days in a year) and divided by the total capital invested (long plus short), which in these backtests is assumed to be $20M ($10M long and $10M short).[79]  The sixth column shows the annualized Sharpe ratio [Sharpe, 1994], which is computed as the average daily P&L divided by

the standard deviation of the daily P&L and multiplied by $\sqrt{252}$ (the annualization factor). The seventh column shows cents-per-share (CPS), which is computed by dividing the total P&L expressed in cents (as opposed to dollars) by the total number of shares traded. The eighth column shows the P&L drawdown (peak-to-trough reduction), between February 20, 2020 (the day on which S&P500 peaked – see above) and March 23, 2020, expressed as a percentage of the total capital invested ($20M).[80] The drawdown applies only to the C2C1, C2C5, C2C10 and C2C20 mean-reversion strategies. The performance characteristics for the D0 strategies, even with trading costs, should be taken with a grain of salt as these are delay-0 signals (see above).

The backtests suggest the following. The close-to-close (C2C) mean-reversion strategies where essentially flatlined (before the costs) even before Quant Bust 2020 (as the market was in the rallying regime – see above). They all had large drawdowns (around 10% for 1-day strategies, around 15% for longer-horizon SIC strategies, and even deeper for longer-horizon STAT strategies).[81] Optimized SIC strategies did better than those based on the weighted regression.[82] The momentum (MOM1) strategies also were essentially flatlined (before the costs) before Quant Bust 2020, and had P&L spikes (opposite to the mean-reversion strategy drawdowns) during Quant Bust 2020 (even after the modeled trading costs). The delay-0 mean-reversion strategies (D0) unsurprisingly had positive returns even before Quant Bust 2020 (and even after the modeled trading costs) and had P&L spikes during Quant Bust 2020 (with the aforesaid "grain-of-salt" comment regarding the D0 strategies). The lesson and the bottom line are that, while the C2C mean-reversion strategies were wobbling around (before the costs) in the upward market prior to Quant Bust 2020, they all had large drawdowns during Quant Bust 2020 for the simple reason we discuss above: the previously existing correlations between the stocks got undone during the COVID-19 selloff, and the mean-reversion strategies reactively got loaded (doubled-down) more and more on losing trades anticipating a snap-back.

---

[80] In our backtests the profits intentionally are not assumed to be reinvested. Since the strategies are simulated to trade purely intraday, it is assumed that a $10M long plus $10M short portfolio is established at each open and the portfolio is liquidated at the close of the same trading day, so P&L from each day is independent from other days.

[81] That fundamental industry classifications (even SIC, which is decent, but not as good a classification as, e.g., GICS or BICS) outperform statistical industry classifications in this context is not news [Kakushadze and Yu, 2016b], [Kakushadze and Yu, 2019]. Well-built fundamental industry classifications are based on analyzing companies' sources of revenues, products, services, customers, partners, suppliers, competitors, etc. (which information can be obtained for publicly traded companies from their 10-Q and 10-K SEC filings and other sources). While parts of this process can be and in many cases are automated, there can be instances where human analysts have to make judgment calls in terms of assigning companies to various industries, sub-industries, etc. Clustering algorithms based on only pricing data cannot compete with the fundamental approach as their signal-to-noise ratio is too low.

[82] We did not include backtests for STAT strategies based on the weighted regression as the results are very similar.

**Table 1.** Backtest results. ROC = annualized return-on-capital (without leverage). Sharpe = annualized Sharpe ratio. CPC = cents-per-share (P&L in cents divided by the total shares traded). Drawdown = peak-to-trough P&L reduction (in %, relative to the total capital invested, $20M) during the period February 20, 2020 to March 23, 2020 (the trough in all cases occurred on March 18, 2020). Costs = whether trading costs are included. OPT/REG = mean-variance optimization or weighted regression. STAT = statistical industry classification. D0 = delay-0 close-to-open return. MOM1 = previous open to previous close return (momentum strategy). C2C{$d$} = $d$-trading-day close-to-close return starting with previous close. See Appendix A.

| Return | Classification | OPT/REG | Costs | ROC | Sharpe | CPC | Drawdown |
|--------|----------------|---------|-------|--------|--------|-------|----------|
| D0 | SIC | REG | N | 27.88 | 4.6 | 1.47 | --- |
| D0 | SIC | REG | Y | 13.79 | 2.28 | 0.74 | --- |
| MOM1 | SIC | REG | N | 10.32 | 1.53 | 0.55 | --- |
| MOM1 | SIC | REG | Y | -5.95 | -0.88 | -0.32 | --- |
| C2C1 | SIC | REG | N | -8.7 | -1.24 | -0.47 | 11.2 |
| C2C1 | SIC | REG | Y | -25.1 | -3.59 | -1.36 | 12.48 |
| C2C5 | SIC | REG | N | -8.5 | -0.99 | -0.45 | 12.97 |
| C2C5 | SIC | REG | Y | -25.48 | -2.95 | -1.35 | 13.94 |
| C2C10 | SIC | REG | N | -10.77 | -1.22 | -0.56 | 14.64 |
| C2C10 | SIC | REG | Y | -27.86 | -3.15 | -1.46 | 15.53 |
| C2C20 | SIC | REG | N | -7.89 | -0.94 | -0.41 | 13.36 |
| C2C20 | SIC | REG | Y | -25.04 | -2.99 | -1.3 | 14.25 |
| D0 | SIC | OPT | N | 31.42 | 6 | 1.81 | --- |
| D0 | SIC | OPT | Y | 18.34 | 3.49 | 1.08 | --- |
| MOM1 | SIC | OPT | N | 8.48 | 1.42 | 0.5 | --- |
| MOM1 | SIC | OPT | Y | -6.5 | -1.09 | -0.39 | --- |
| C2C1 | SIC | OPT | N | -8.48 | -1.39 | -0.5 | 9.92 |
| C2C1 | SIC | OPT | Y | -23.57 | -3.85 | -1.41 | 10.9 |
| C2C5 | SIC | OPT | N | -8.64 | -0.96 | -0.51 | 13.9 |
| C2C5 | SIC | OPT | Y | -24.29 | -2.71 | -1.43 | 14.7 |
| C2C10 | SIC | OPT | N | -10.9 | -1.21 | -0.63 | 15.64 |
| C2C10 | SIC | OPT | Y | -26.64 | -2.96 | -1.56 | 16.45 |
| C2C20 | SIC | OPT | N | -8.33 | -0.99 | -0.48 | 14.13 |
| C2C20 | SIC | OPT | Y | -24.02 | -2.85 | -1.4 | 14.95 |
| D0 | STAT | OPT | N | 24.73 | 4.77 | 1.5 | --- |
| D0 | STAT | OPT | Y | 12.92 | 2.4 | 0.8 | --- |
| MOM1 | STAT | OPT | N | 7.02 | 1.25 | 0.44 | --- |
| MOM1 | STAT | OPT | Y | -7.12 | -1.11 | -0.45 | --- |
| C2C1 | STAT | OPT | N | -7.42 | -1.24 | -0.46 | 9.91 |
| C2C1 | STAT | OPT | Y | -21.03 | -3.65 | -1.32 | 10.84 |
| C2C5 | STAT | OPT | N | -10.55 | -1.1 | -0.64 | 16.39 |
| C2C5 | STAT | OPT | Y | -24.8 | -2.62 | -1.52 | 16.81 |
| C2C10 | STAT | OPT | N | -11.63 | -1.21 | -0.7 | 16.95 |
| C2C10 | STAT | OPT | Y | -27.13 | -2.82 | -1.65 | 18.08 |
| C2C20 | STAT | OPT | N | -11.21 | -1.18 | -0.67 | 16.95 |
| C2C20 | STAT | OPT | Y | -26.44 | -2.84 | -1.59 | 17.47 |

**Figure 1.** P&L plots over a 1-year period ending March 23, 2020 for the strategies with the following attributes: SIC, REG, N (columns 2-4 of Table 1); ordered first left-to-right and then top-to-bottom in the graph: C2C20, C2C10, C2C5, C2C1, MOM1, D0 (in column 1 of Table 1).

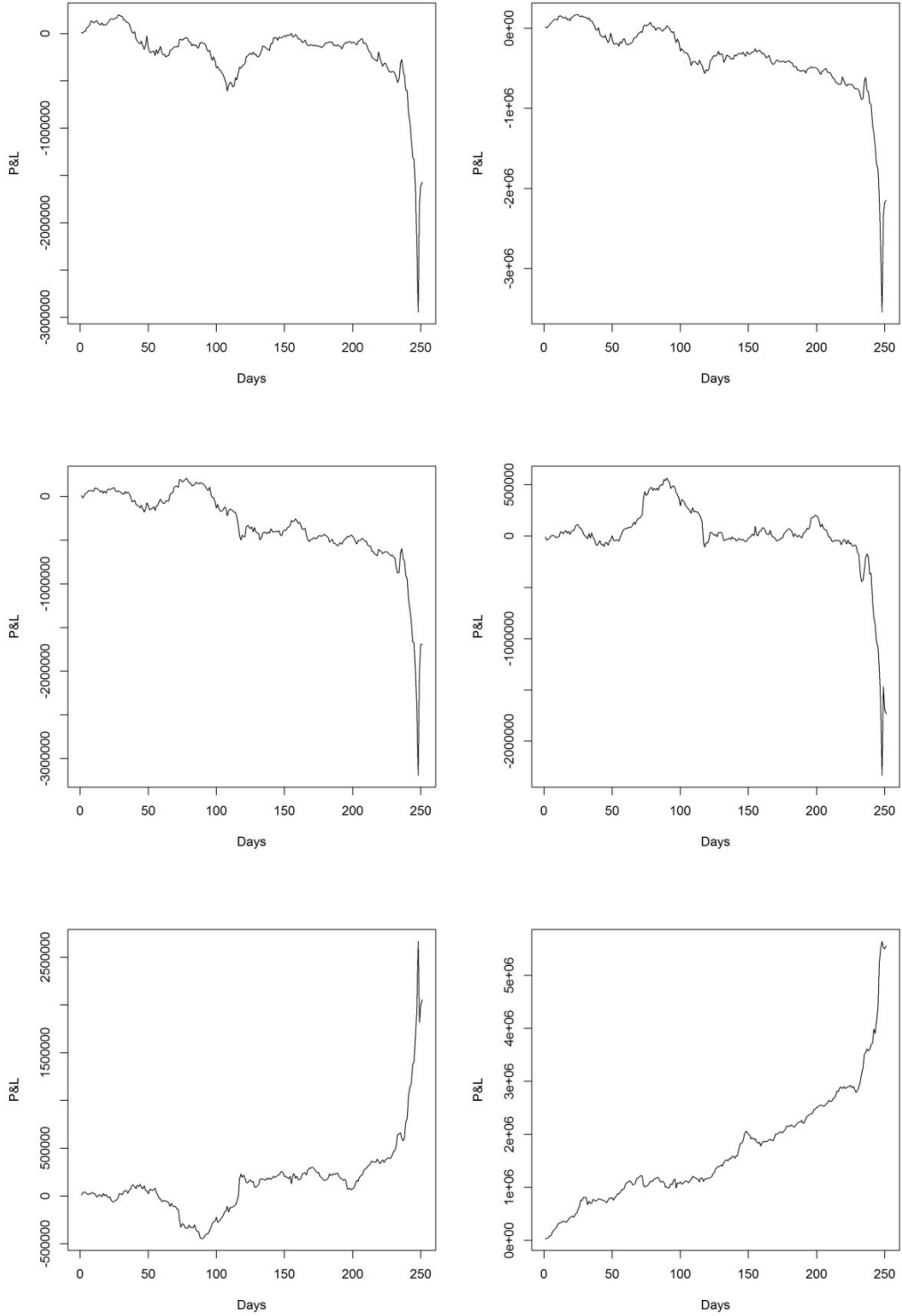

**Figure 2.** P&L plots over a 1-year period ending March 23, 2020 for the strategies with the following attributes: SIC, REG, Y (columns 2-4 of Table 1); ordered first left-to-right and then top-to-bottom in the graph: C2C20, C2C10, C2C5, C2C1, MOM1, D0 (in column 1 of Table 1).

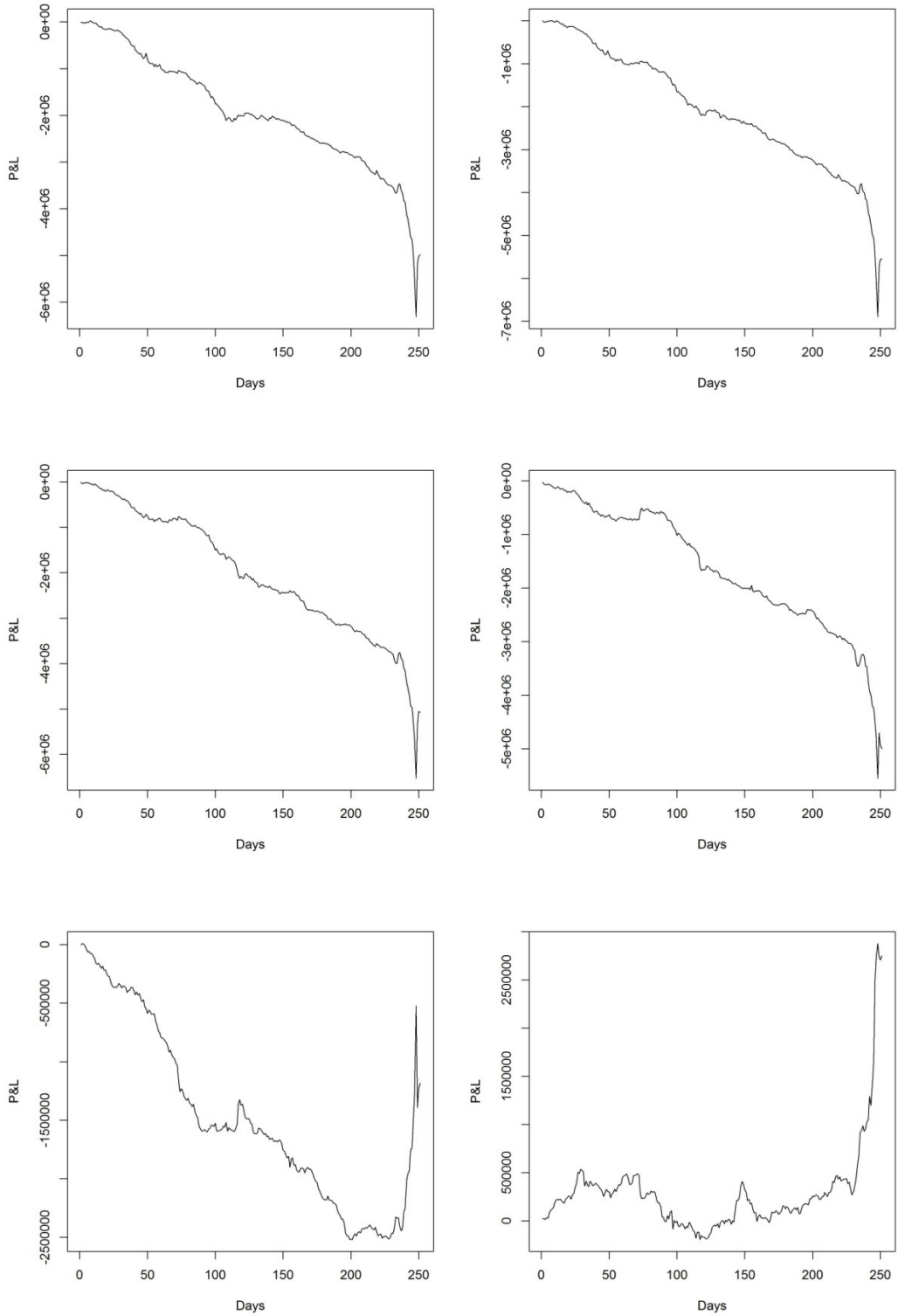

**Figure 3.** P&L plots over a 1-year period ending March 23, 2020 for the strategies with the following attributes: SIC, OPT, N (columns 2-4 of Table 1); ordered first left-to-right and then top-to-bottom in the graph: C2C20, C2C10, C2C5, C2C1, MOM1, D0 (in column 1 of Table 1).

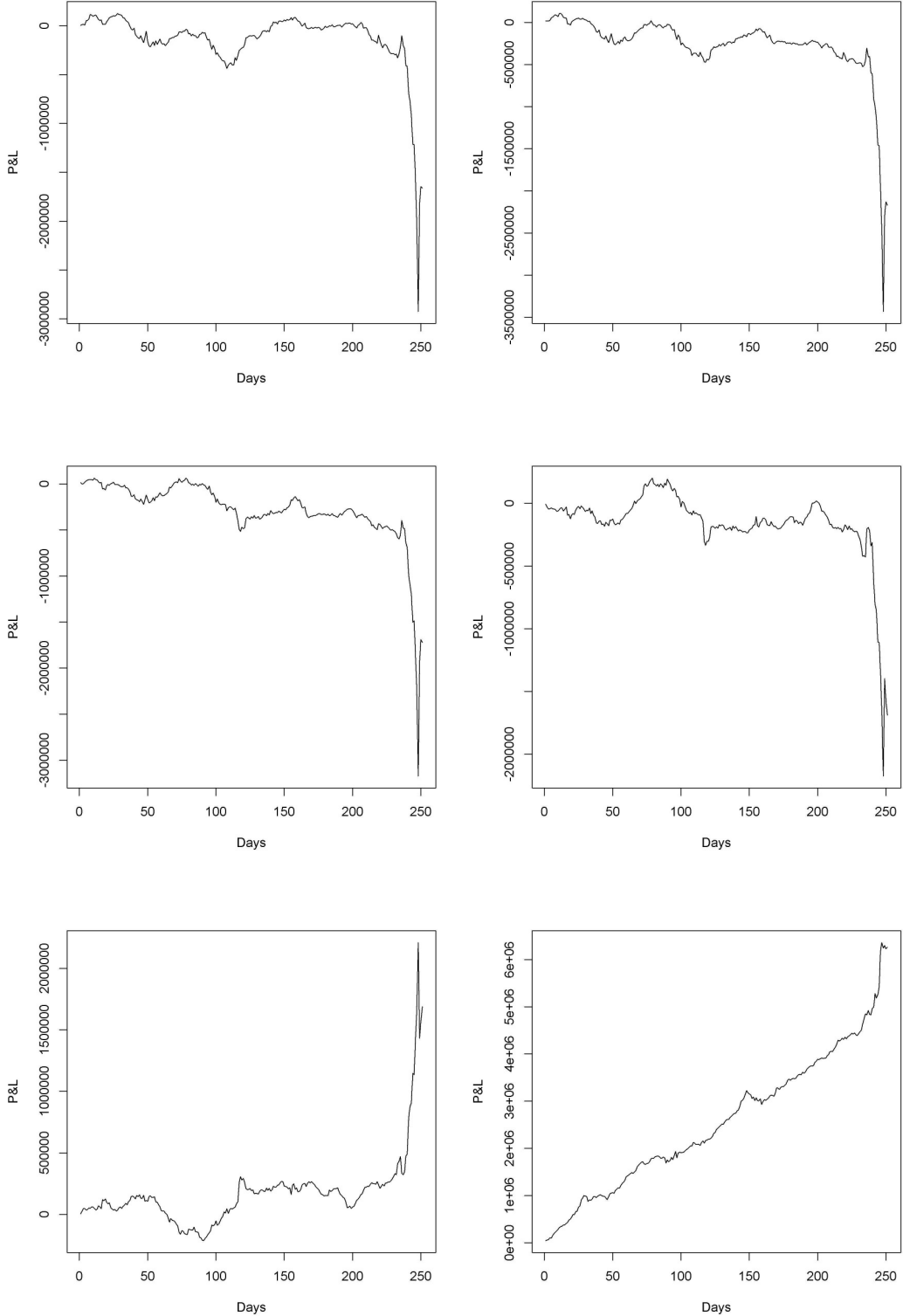

**Figure 4.** P&L plots over a 1-year period ending March 23, 2020 for the strategies with the following attributes: SIC, OPT, Y (columns 2-4 of Table 1); ordered first left-to-right and then top-to-bottom in the graph: C2C20, C2C10, C2C5, C2C1, MOM1, D0 (in column 1 of Table 1).

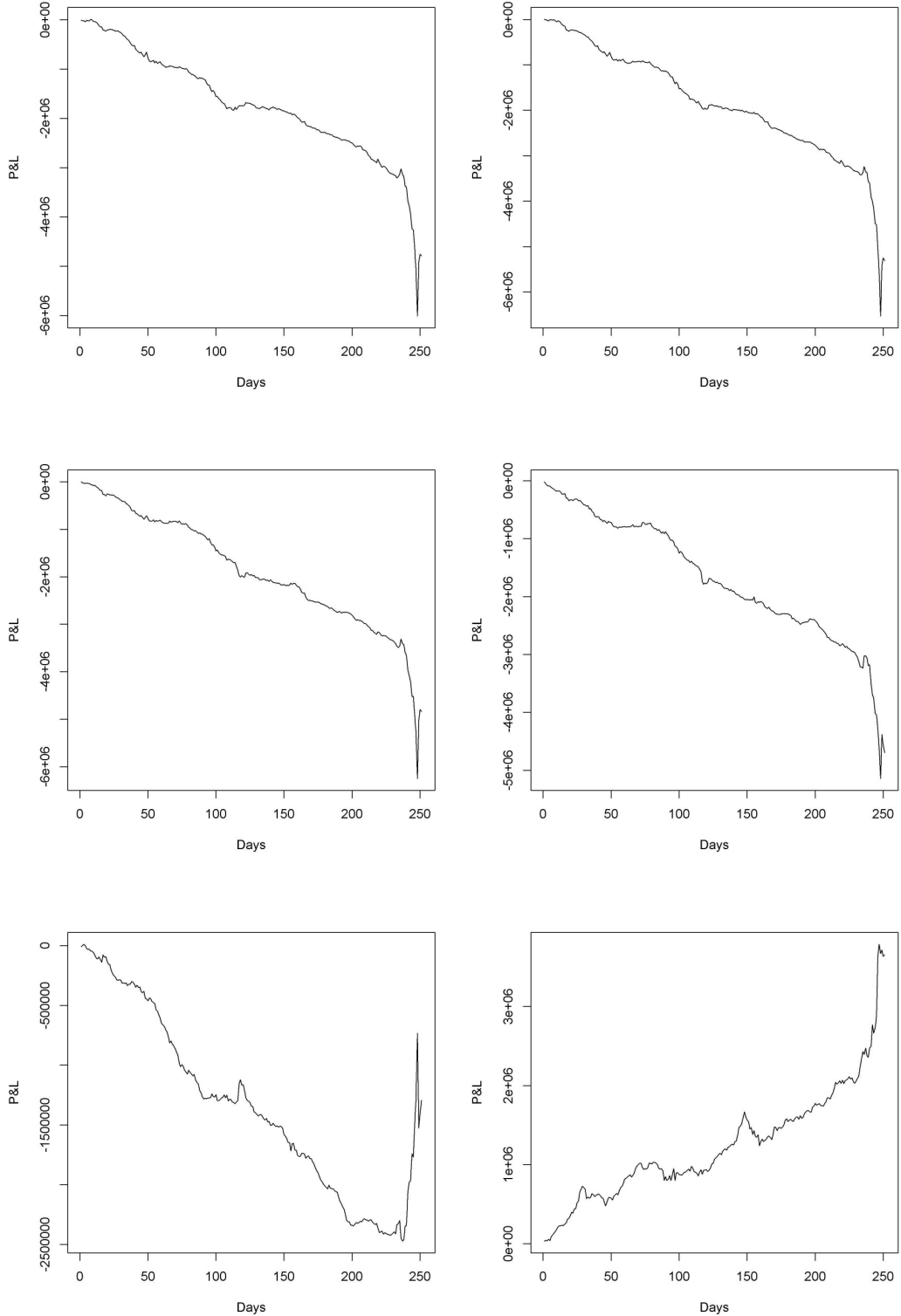

**Figure 5.** P&L plots over a 1-year period ending March 23, 2020 for the strategies with the following attributes: STAT, OPT, N (columns 2-4 of Table 1); ordered first left-to-right and then top-to-bottom in the graph: C2C20, C2C10, C2C5, C2C1, MOM1, D0 (in column 1 of Table 1).

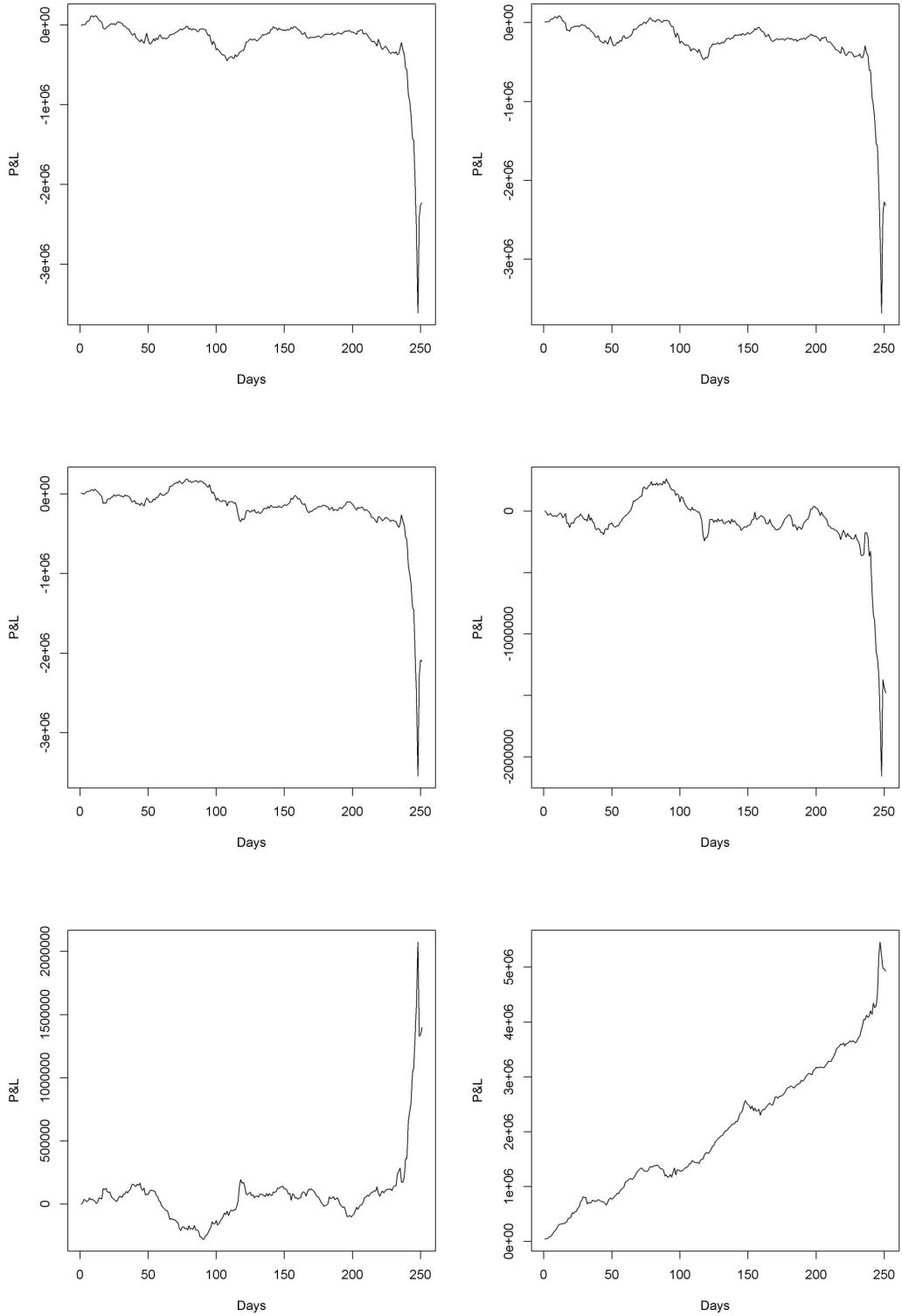

**Figure 6.** P&L plots over a 1-year period ending March 23, 2020 for the strategies with the following attributes: STAT, OPT, Y (columns 2-4 of Table 1); ordered first left-to-right and then top-to-bottom in the graph: C2C20, C2C10, C2C5, C2C1, MOM1, D0 (in column 1 of Table 1).

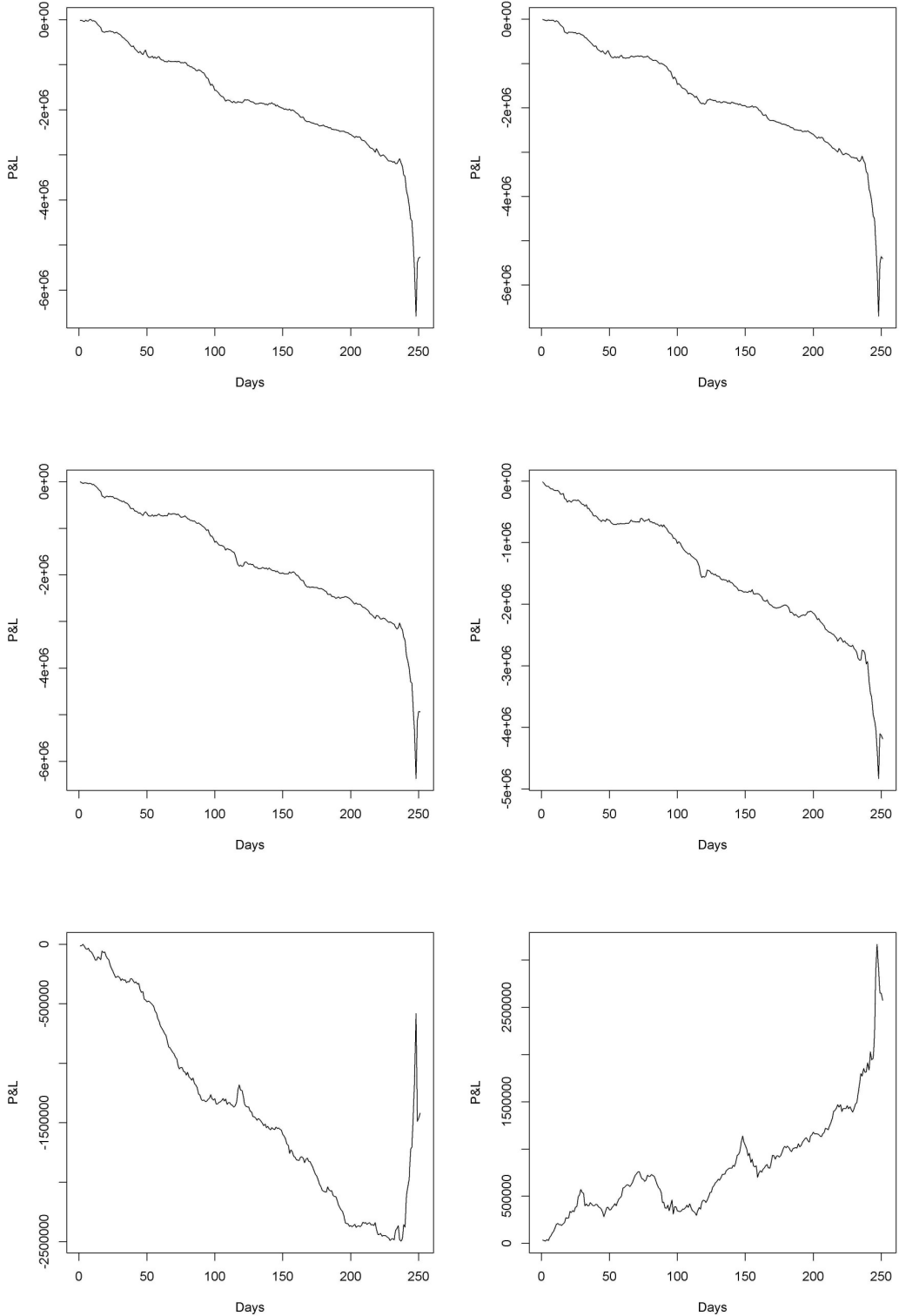